\documentclass[twocolumn]{aastex62}

\usepackage{multirow,arydshln,hhline}

\submitjournal{ApJ}

\shorttitle{OGLE RR Lyrae Stars in the Magellanic Bridge}
\shortauthors{Jacyszyn-Dobrzeniecka et al.}

\begin{document}

\title{OGLE-ING THE MAGELLANIC SYSTEM: RR LYRAE STARS IN THE BRIDGE\footnote{Draft version prepared on April, 16th, 2019}}

\correspondingauthor{Anna M. Jacyszyn-Dobrzeniecka}
\email{jacyszyn@uni-heidelberg.de}

\author[0000-0002-5649-536X]{Anna M. Jacyszyn-Dobrzeniecka}
\affiliation{Astronomical Observatory, University of Warsaw, Al. Ujazdowskie 4, 00-478 Warszawa, Poland}
\affiliation{Astronomisches Rechen-Institut, Zentrum f\"ur Astronomie der Universit\"at Heidelberg, M\"onchhofstr. 12-14, D-69120 Heidelberg, Germany}

\author{Przemek Mr\'oz}
\affiliation{Astronomical Observatory, University of Warsaw, Al. Ujazdowskie 4, 00-478 Warszawa, Poland}

\author{Katarzyna Kruszy\'nska}
\affiliation{Astronomical Observatory, University of Warsaw, Al. Ujazdowskie 4, 00-478 Warszawa, Poland}

\author{Igor Soszy\'nski}
\affiliation{Astronomical Observatory, University of Warsaw, Al. Ujazdowskie 4, 00-478 Warszawa, Poland}

\author{Dorota M. Skowron}
\affiliation{Astronomical Observatory, University of Warsaw, Al. Ujazdowskie 4, 00-478 Warszawa, Poland}

\author{Andrzej Udalski}
\affiliation{Astronomical Observatory, University of Warsaw, Al. Ujazdowskie 4, 00-478 Warszawa, Poland}

\author{Micha\l{} K. Szyma\'nski}
\affiliation{Astronomical Observatory, University of Warsaw, Al. Ujazdowskie 4, 00-478 Warszawa, Poland}

\author{Patryk Iwanek}
\affiliation{Astronomical Observatory, University of Warsaw, Al. Ujazdowskie 4, 00-478 Warszawa, Poland}

\author{Jan Skowron}
\affiliation{Astronomical Observatory, University of Warsaw, Al. Ujazdowskie 4, 00-478 Warszawa, Poland}

\author{Pawe\l{} Pietrukowicz}
\affiliation{Astronomical Observatory, University of Warsaw, Al. Ujazdowskie 4, 00-478 Warszawa, Poland}

\author[0000-0002-9245-6368]{Rados\l{}aw Poleski}
\affiliation{Department of Astronomy, Ohio State University, 140 West 18th Avenue, Columbus, OH 43210, USA}

\author{Szymon Koz\l{}owski}
\affiliation{Astronomical Observatory, University of Warsaw, Al. Ujazdowskie 4, 00-478 Warszawa, Poland}

\author{Krzysztof Ulaczyk}
\affiliation{Department of Physics, University of Warwick, Coventry CV4 7AL, UK}

\author{Krzysztof Rybicki}
\affiliation{Astronomical Observatory, University of Warsaw, Al. Ujazdowskie 4, 00-478 Warszawa, Poland}

\author{Marcin Wrona}
\affiliation{Astronomical Observatory, University of Warsaw, Al. Ujazdowskie 4, 00-478 Warszawa, Poland}

\begin{abstract}

We use the extended and updated OGLE Collection of Variable Stars to thoroughly analyze distribution of RR Lyrae stars in the Magellanic Bridge. We use photometric metallicities to derive absolute Wesenheit magnitude and then individual distance of each RR Lyr star. We confirm results from our earlier study showing that RR Lyr stars are present in between the Magellanic Clouds, though their three-dimensional distribution rather resembles two extended overlapping structures than a strict bridge-like connection. The contours do connect in the southern parts of the Bridge, albeit on a too low level to state that there exists an evident connection. To test the sample numerically, we use multi-Gaussian fitting and conclude that there is no additional population or overdensity located in the Bridge. We also try to reproduce results on putative RR Lyr Magellanic Bridge stream by selecting RR Lyr candidates from {\it Gaia} Data Release 1. We show that we are not able to obtain the evident connection of the Clouds without many spurious sources in the sample, as the cuts are not able to remove artifacts and not eliminate the evident connection at the same time. Moreover, for the first time we present the {\it Gaia} Data Release 2 RR Lyr stars in the Magellanic Bridge area and show that their distribution matches our results.

\end{abstract}


% % % % % %

\section{Introduction} \label{sec:intro}

Interactions between the Magellanic Clouds, and probably between the pair and the Milky Way, led to a formation of an entire complex of structures, together with the Clouds referred to as the Magellanic System (e.g. \citealt{Gardiner1994,Gardiner1996,Yoshizawa2003,Connors2006,Ruzicka2009,Ruzicka2010,Besla2010,Besla2012,Diaz2011,Diaz2012,Guglielmo2014}). One of direct evidences of the latest encounter of the Large and Small Magellanic Clouds (LMC and SMC, respectively) is the Magellanic Bridge (MBR; i.e. \citealt{Harris2007}).

Many studies proved that there are young stars located in between the LMC and SMC \citep{Shapley1940,Irwin1985,Demers1998,Harris2007,Noel2013,Noel2015}, and moreover, that they form a continuous connection matching the neutral hydrogen (\textsc{H i}) contours \citep{Skowron2014}. Young ages of some objects suggest an \textit{in-situ} Bridge formation (e.g. Jacyszyn-Dobrzeniecka et al. 2016, 2019, hereafter \citealt{PaperI} and \citealt{PaperIII}, respectively). This implies that the interactions were strong enough to pull out gas from the Magellanic Clouds and trigger star formation outside these galaxies. For better understanding of processes leading to these events, it is also important to test the older stellar populations in the MBR. Were the interactions strong enough to pull out not only gas, but also stars from either or both LMC and SMC? Hereafter we focus on the older population of stars. For more information about different characteristics of the Bridge see introduction in \citealt{PaperIII}.

Candidates for a stellar Bridge counterpart belonging to the older population were found by \citet{Bagheri2013} and \citet{Skowron2014}. \citet{WagnerKaiser2017} analyzed RR Lyrae (RRL) stars using the Optical Gravitational Lensing Experiment (OGLE) Collection of RRL stars and demonstrated that there exists a continuous flow of these objects between the Magellanic Clouds. Authors point out that metallicities and distances of old population members show a smooth transition between the LMC and SMC. Moreover, the RRL stars distribution is not matching the \textsc{H i} density distribution. Thus, they suggest that RRL stars are rather resembling two overlapping structures than a tidally stripped bridge. Recently, \citet{Zivick2018} used {\it Gaia} data to show that old stellar population is more broadly distributed and does not follow the \textsc{H i} bridge, in contrary to young population.

Jacyszyn-Dobrzeniecka et al. (2017, hereafter \citealt{PaperII}) used the same sample from the OGLE Collection of Variable Stars (OCVS, \citealt{Soszynski2016}) to analyze three-dimensional distribution of RRL stars in the Magellanic System and the Bridge as well. Their results are perfectly consistent with those of \citet{WagnerKaiser2017}, showing that there is not much evidence for a bridge-like structure formed by old population between the Magellanic Clouds.

On the other hand, \citet{Carrera2017} studied 39 intermediate-age and old stars in two Bridge fields located near highest \textsc{H i} density contours and close to the SMC (between RA $2^{\rm h}$ and $3^{\rm h}$) and found that, based on chemistry and kinematics, these objects are tidally stripped from the SMC. Their metallicities are consistent with those of \citet{WagnerKaiser2017}. Both results are not necessarily incoherent, as stars analyzed by \citet{Carrera2017} may just be SMC halo members. The kinematics are in agreement with recent studies by \citet{Oey2018} and \citet{Zivick2018} who found that both young and old stellar populations are moving away from the SMC toward the LMC.

Another study of the Bridge old population was carried out using {\it Gaia} Data Release 1 (DR1, \citealt{Gaia2016}). Belokurov et al. (2017, hereafter \citealt{B17}) developed a procedure to select RRL candidates from DR1 and analyzed their distribution in the MBR. They found an evident stellar bridge between the Magellanic Clouds, which is shifted from the young stars bridge, and thus from the highest \textsc{H i} density contours, by about $5\arcdeg$. \citealt{B17} explain this difference by an older bridge trailing rather than following the Magellanic System. Moreover, they also perform a simulation to test whether such scenario is plausible. Later, at least one stellar substructure partially co-spatial with the \citealt{B17} RRL bridge was found by \citet{Mackey2018}, who used deep, panoramic survey conducted with Dark Energy Camera. Also, \citet{Belokurov2018} found such substructures in red giants distribution using {\it Gaia} Data Release 2 (DR2).

Similarly to \citealt{B17}, \citet{Deason2017} selected Mira candidates from DR1 and analyzed their distribution in the Magellanic System. They found that there are not as many Miras as RRL stars in the Bridge and no bridge-like connection could be found. However, Miras form a slightly extended feature stretching out of the SMC toward the RRL bridge discovered by \citealt{B17}.

In this paper, which is the fourth in the series devoted to analysis of three-dimensional structure of the Magellanic System using the OCVS, we examine the RRL stars distribution in the Bridge area with an extended and updated OGLE data. We also compare our results to \citealt{B17}, whose results are not in agreement with \citealt{PaperII}. Moreover, we perform an analysis of the DR1 data using \citealt{B17} method and show their distribution of RRL candidates. We also show for the first time the distribution of RRL stars from the {\it Gaia} DR2 (\citealt{Gaia2018,Holl2018,Clementini2019}) in the Bridge area.

We organized the paper as follows. Section~2 describes the RRL stars from the OCVS and the updates, corrections and extensions that were lately applied to the Collection. Samples selection as well as methods used for analysis, are found in Section~3. In Section~4 we describe a study of three-dimensional distribution of RRL stars from the OCVS. Section~5 presents a reanalysis of OCVS sample using different method, which is an attempt to reproduce \citealt{B17} results. In Section~6 we present our analysis of DR1 data using \citealt{B17} method to select RRL candidates. Section~7 presents DR2 RRL stars distribution in the MBR. In Section~8 we compare distributions of different stellar tracers in the Bridge. We conclude the paper in Section~9.

% % % % % %

\section{Observational Data}

\begin{figure*}[htb]
	\includegraphics[width=\textwidth]{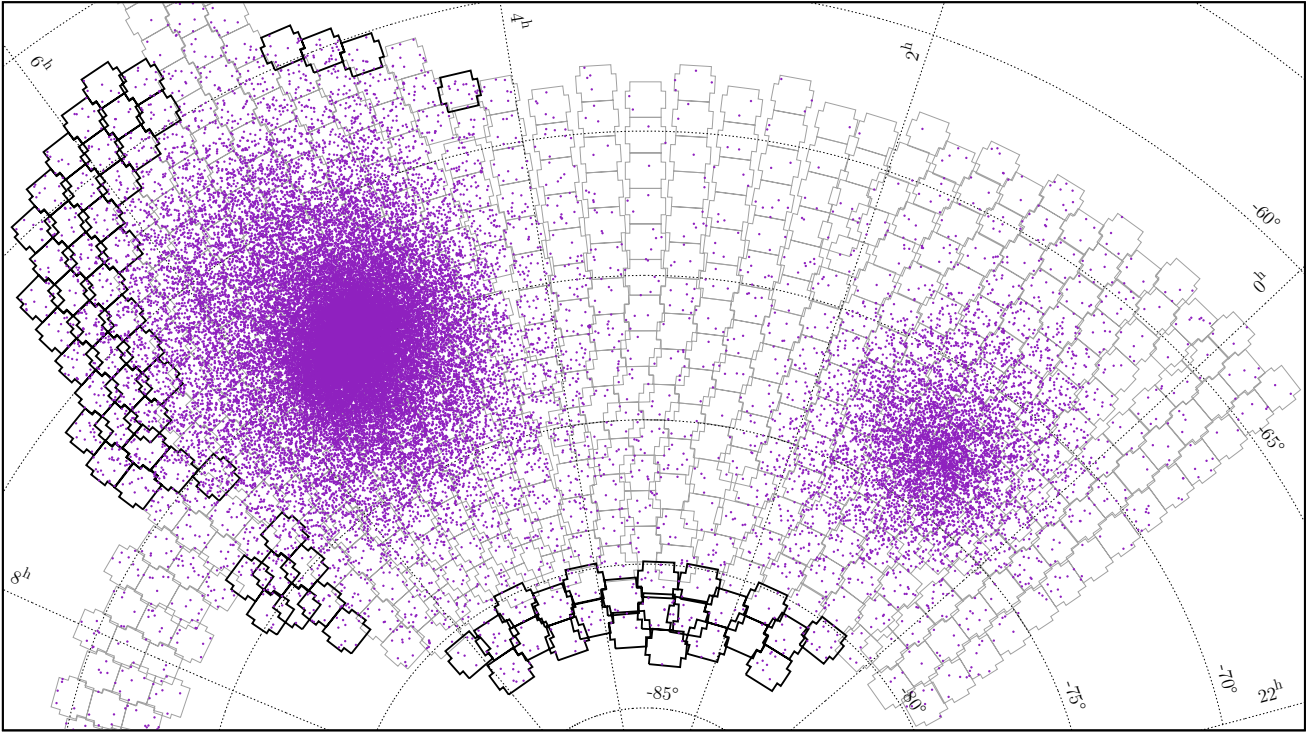}
	\caption{The on-sky locations of RRL stars in the Magellanic System. Black contours show the newest addition to the OGLE-IV fields while grey show main OGLE-IV fields in the Magellanic System that were already observed before July 2017.}
	\label{fig:rrl-all}
\end{figure*}

\subsection{OGLE Collection of Variable Stars} \label{sec:ocvs}

Since \citealt{PaperII} was published, the OCVS was already updated and a number of new RRL pulsators were added \citep{Soszynski2016,Soszynski2017}. In this paper, similarly to \citealt{PaperIII}, we use the newest data from the OCVS. For more technical details about the fourth phase of the OGLE project see \citet{Udalski2015}.

In \citealt{PaperIII} we presented the latest updates that were applied to the OGLE Collection of Cepheids since \citet{Soszynski2017} and were not yet published. Similar changes also affect the RRL stars Collection. The newest version that we use here includes 1242 RRL stars that were added to the sample. Tab.~\ref{tab:rrl-add} presents exact numbers of RRL stars of different types that were added from different sources. The largest number of newly included objects was added from the newest fields located east and south of the LMC -- almost a thousand of RRL stars. The newest fields in the southern parts of the Magellanic Bridge resulted in an addition of over 100 of RRL pulsators. For a current OGLE-IV footprint with the newly added fields and the on-sky distribution of all OCVS RRL stars see Fig.~\ref{fig:rrl-all}.

\setlength\dashlinedash{1pt}
\setlength\dashlinegap{2pt}

\begin{table}[htb]
\caption{Latest additions to the OGLE Collection of RRL stars}
\label{tab:rrl-add}
\centering
\begin{tabular}{lDDDD}

Source & \multicolumn2c{RRab} & \multicolumn2c{RRc} & \multicolumn2c{RRd} & \multicolumn2c{All} \\
%\hline
\cline{1-9}
\decimals
New MBR fields$^{(a)}$   & 102 &  40 & 10 & 152 \\
New LMC fields$^{(a)}$   & 679 & 234 & 71 & 984 \\
{\it Gaia} DR2$^{(b)}$         &  98 &   7 &  1 & 106 \\
\cdashline{1-9}
All                      & 879 & 281 & 82 & 1242 \\
%\hline
\cline{1-9}
\multicolumn{9}{p{.38\textwidth}}{$(a)$ For a current OGLE-IV footprint with these newly added fields see Fig.~1 in \citealt{PaperIII}. $(b)$ We searched for RRL stars that are present in {\it Gaia} DR2 but not in OCVS. After careful studies of their OGLE lightcurves we classified a number of additional RRL stars in the Magellanic System.}

\end{tabular}
\end{table}

We also cross-matched the {\it Gaia} DR2 RRL stars \citep{Gaia2018,Holl2018,Clementini2019} with OCVS and carefully searched for any RRL pulsators that were not present in the OGLE database. Based on their OGLE lightcurves we additionally classified 106 RRL stars.

% % % % % %

\section{Data Analysis}

\subsection{Sample Selection}

In our basic approach we use a very similar method to \citealt{PaperII} and this technique is different to the one we were using for the Cepheid sample in \citealt{PaperIII}. Hereafter, we only analyse the RRab stars as these are the most common type and about 70$\%$ of all RRL stars pulsate solely in the fundamental mode (i.e. see number of RRL stars published by \citealt{Soszynski2011,Soszynski2014,Soszynski2016,Soszynski2017}).

We select a few different samples from the entire OGLE Collection of RRL stars in the Magellanic System. The first sample (hereafter the entire sample) contains all of the RRab stars and can be only represented in the on-sky maps. All of the RRab stars for which we were able to calculate distance constitute the second sample (hereafter the uncleaned sample). These stars must have both $I$- and $V$-passband magnitudes and a well estimated $\phi_{31}$ coefficient (this is one of the lightcurve Fourier decomposition parameters, \citealt{Simon1981}). To create the third sample (the cleaned sample) we made an additional cut on the Bailey diagram, the same as we did in \citealt{PaperII} (see Sec.~2.2 and Fig.~1 therein for more details). Then we fitted PL relations to the second sample using Wesenheit magnitude and iteratively applied $3\sigma$ clipping to the data after each fit (see Sec.~3.1 in \citealt{PaperII} for more details). Any other additional cuts or selections made to the three described samples are discussed later.

Taking into account the updates made and less complicated cleaning process this sample should not be fully consistent with our \citealt{PaperII} sample.

\subsection{Individual Distances and Coordinates}

To calculate individual distances of RRab stars we use exactly the same method as we did in \citealt{PaperII} and in \citet{Skowron2016}. For the determination of photometric metallicity $\phi_{31}$, that we obtained from Fourier decomposition of OGLE lightcurves, we apply relation from \citet{Nemec2013}. Then we use relations from \citet{Braga2015} to calculate absolute Wesenheit magnitudes. Having these values and the observed magnitudes, we were able to determine distance to each RRab star. For more details on used relations and exact transformations see Sec.~3.2 in \citealt{PaperII} and Sec.~5 in \citet{Skowron2016}.

Similarly to \citealt{PaperI}, \citealt{PaperII} and \citealt{PaperIII}, we use a Hammer-equal area projection for on-sky plots and Cartesian three-dimensional coordinate system. The exact equations can be found in Sec.~3.2 of \citealt{PaperIII} (Eqs.~$1-5$).

% % % % % %

\section{OGLE RRL Sample}

\subsection{Three-Dimensional Distribution}

\begin{figure*}[htb]
	\includegraphics[width=\textwidth]{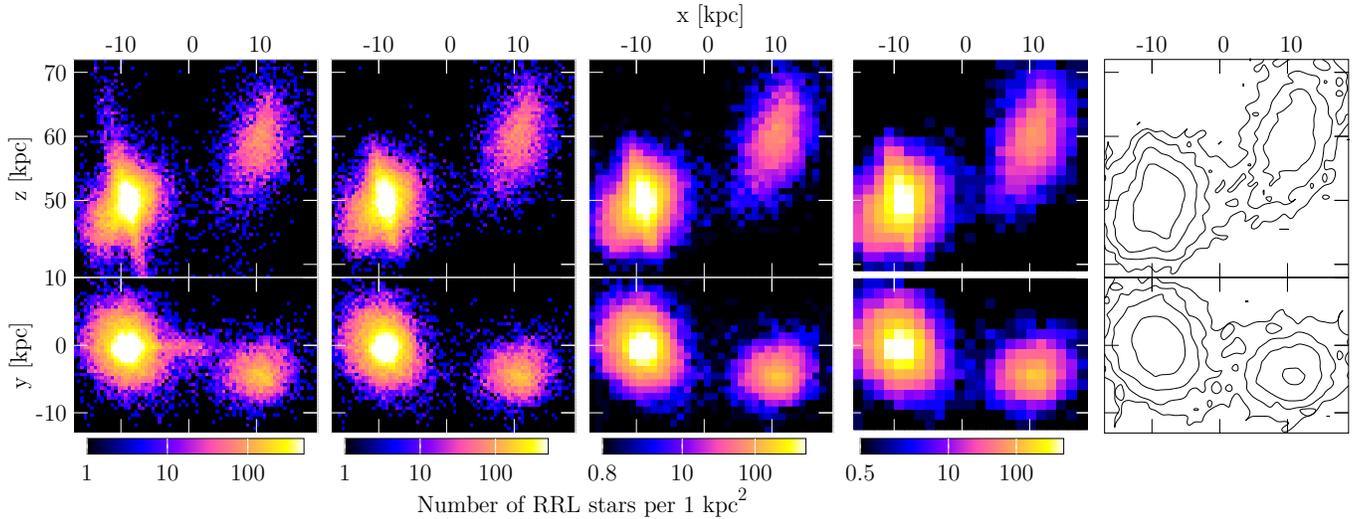}
	\caption{Top (upper row) and front (bottom row) view of the RRab stars in the Magellanic System using Cartesian space projections. Left panels show the uncleaned sample. The ''blend-artifact'', a non-physical structure seemingly emanating from the LMC center, is very clearly visible. It disappears on the other panels, where we show the cleaned sample. Three middle panels present the same sample but with different bin sizes -- from left to right -- 0.5, 1 and 1.5 kpc. The right panels show contours fitted to the middle panels. Contours are on the levels of 1, 5, 20, 100 RRab stars per kpc$^2$. The lines do connect but on a very low level.}
	\label{fig:rrl-cart}
\end{figure*}

Fig.~\ref{fig:rrl-cart} shows top (upper row) and front (bottom row) view of the three-dimensional RRab stars distribution in the Magellanic System. The plots were made using two-dimensional Cartesian space projections. Left panels show the uncleaned sample with a very clearly visible ''blend-artifact'' in the LMC. This is a non-physical structure that seems to be emanating from the LMC center and is caused by blending and crowding effects (for more detailed description see Sec.~2.2 and Fig.~3 in \citealt{PaperII}). The ''blend-artifact'' is not that clearly visible in the next panels, where we show the cleaned sample. Three middle panels show the same sample but with different bin sizes. Contours fitted to the middle panels are shown in the right panels. The lines are on the levels of 1, 5, 20, 100 RRab stars per kpc$^2$.

All of the panels in Fig.~\ref{fig:rrl-cart} show the Bridge area. As in \citealt{PaperII}, we do see some RRab stars located between the Magellanic Clouds. These objects may be forming halos, though some evidence was found that the LMC may have as well an extended disk \citep{Saha2010,Balbinot2015,Besla2016,Mackey2016,Nidever2018}. However, again, we do not see any evident bridge-like connection between the Magellanic Clouds formed by RRL stars in any dimension -- neither $xz$ nor $xy$ projection. Note that the $xy$ projection is very similar to the on-sky view. The contours do connect but on a very low level (1 star per 1~kpc$^2$ and below). It is too low to state, based on the maps only, that there is an overdensity or an evident connection in the Bridge area. Based on three-dimensional maps we can only state that we do see two extended structures overlapping.

% %
\subsection{Numerical Analysis}

To analyse our RRab sample numerically, we performed the multi-Gaussian fitting to our cleaned sample. We approximate the spatial distribution using a Gaussian mixture model with 32 components. The underlying space density of stars is approximated as a sum of Gaussians. Their relative weights and parameters (means, covariances) are found using an expectation-maximization algorithm \citep{Dempster1977} implemented in the Python scikit-learn package \citep{Pedregosa2011}. We tested whether the multi-Gaussian fitting properly describes our data by comparing histograms of real distribution of stars with the simulated ones. We did not specify any parameters -- only the number of Gaussians and the three-dimensional locations of stars from our sample. We tested separately models with 32, 64, 128 and 256 Gaussians and did not find any significant difference between obtained results.

\begin{figure}[htb]
	\begin{center}
	\includegraphics[width=.47\textwidth]{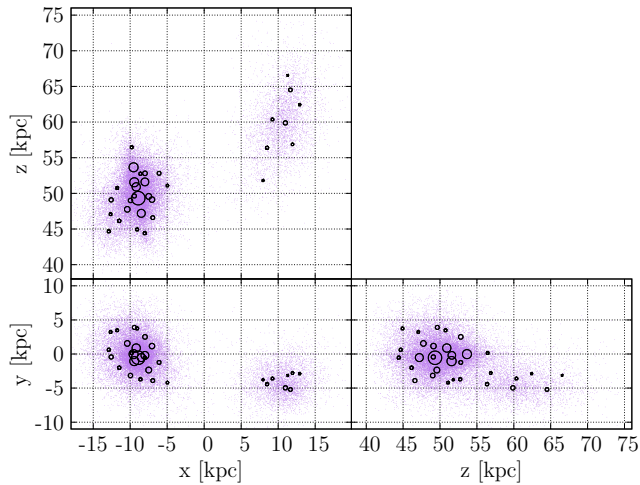}
	\end{center}
	\caption{Centers of 32 fitted Gaussians overplotted on the three-dimensional distribution of RRab stars from the cleaned sample to which the fit was performed. Each Gaussian center is represented as an open circle while the circle size marks number of stars included in each Gaussian. No Gaussian is centered in the genuine Bridge area leading to a conclusion that there is no additional population or overdensity located there.}
	\label{fig:rrl-mod}
\end{figure}

Results of the multi-Gaussian procedure for 32 Gaussians are shown in Fig.~\ref{fig:rrl-mod} where we overplotted Gaussian centers on the three-dimensional distribution of RRab stars from our sample. Each resulting Gaussian is represented with an open circle. The circle size marks number of stars included in each Gaussian: the smallest circle represents 237 objects, the largest -- 2362 objects. The circle radius increases linearly with the number of objects.

Fig.~\ref{fig:rrl-mod} shows that all of the Gaussians are centered in either LMC or SMC and none of them is centered in the genuine Bridge area. This leads to a conclusion that there is no additional population or overdensity located there. Note that this does not mean that there are no stars in the Bridge as the Gaussians have their own individual spread. The Bridge RRab stars are thus modelled as objects located in the Gaussians wings.

\begin{figure}[htb]
	\begin{center}
	\includegraphics[width=.47\textwidth]{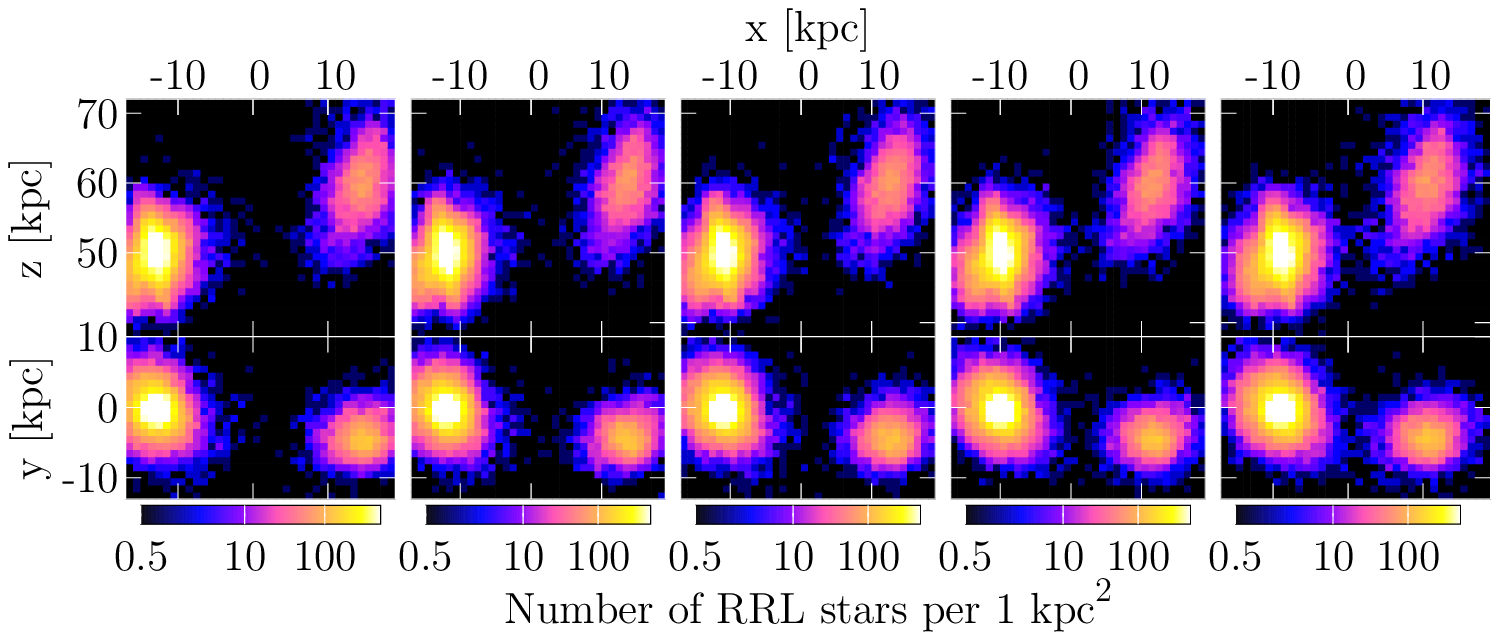} \\
	\includegraphics[width=.47\textwidth]{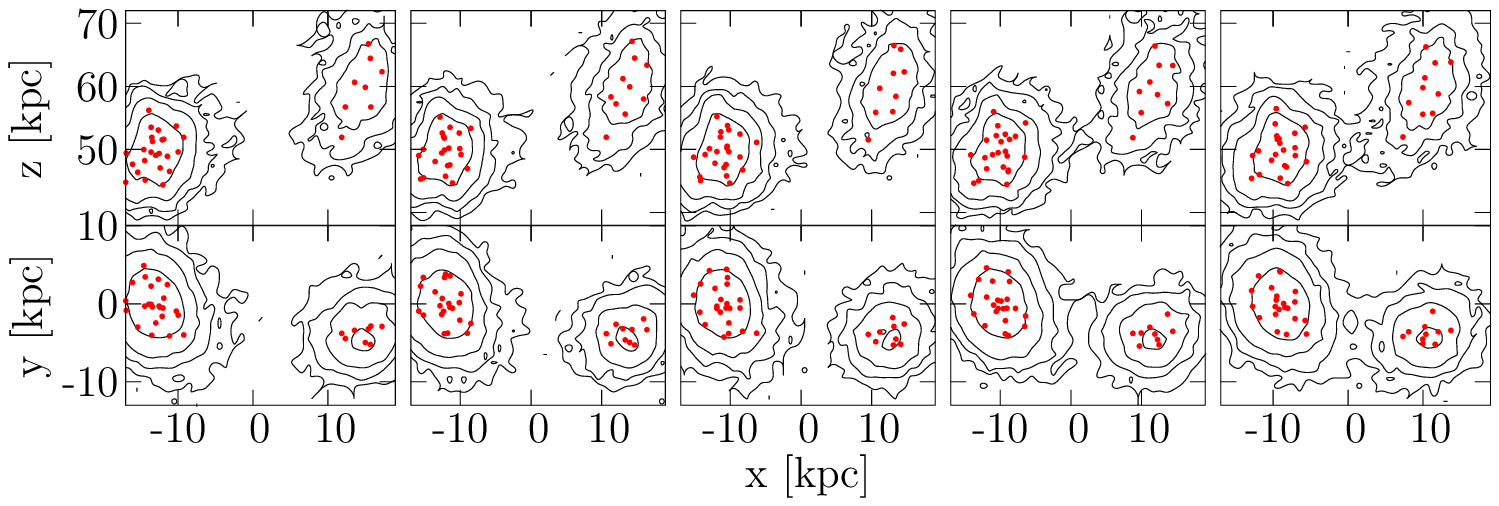}
	\end{center}
	\caption{Two-dimensional plots of three-dimensional Cartesian space projections showing points simulated using multi-Gaussian fit. Top panel shows binned data while bottom -- fitted contours (black lines) and Gaussians centers (red points). Each column represents different separation between LMC and SMC samples starting with 8 kpc in the left column and decreasing by 2 kpc toward right. Right column shows points simulated for no additional offset. The bin size is 1~kpc along every axis and the color scale is the same on each plot in the top panel. The contours are on the same levels as in Fig.~\ref{fig:rrl-cart}, namely 1, 5, 20 and 100 RRab stars per 1~kpc$^2$.}
	\label{fig:rrl-sep}
\end{figure}

To show how and when the contours connect we used multi-Gaussian fit to simulate distribution of objects in the Magellanic System while adding an offset to each Magellanic Cloud sample. We use three-dimensional Cartesian coordinates of our cleaned sample and add an offset to the $x$ coordinate of each Magellanic Cloud sample -- separately for the LMC and SMC. We then fit the Gaussians and simulate locations of the exact number of points that is included in our cleaned sample, precisely 27 212. We bin the data and fit contours. The results are shown in Fig.~\ref{fig:rrl-sep}. The top panel shows binned data with color-coded column density while the bottom panel shows contours (black lines) and Gaussian centers (red points). The bin size and contour levels are the same as in middle and right columns of Fig.~\ref{fig:rrl-cart}. The total offset added or subtracted from the $x$ coordinate decreases from left to right. In the left column, the offset is 8 kpc (4 kpc added in the case of SMC, 4 kpc subtracted for the LMC) and decreases by 2 kpc in each column. Right column shows simulated data with no additional offset. Comparing this column to the middle column of Fig.~\ref{fig:rrl-cart} it is clearly visible that the multi-Gaussian fit reconstructs the real three-dimensional distribution of our data very well.

In the left column, where the distance between the LMC and SMC is the largest, the contours do not connect and these galaxies are separated. Once we reduce the offset, the lowest contours finally connect at the level of 2 kpc of additional offset. The galaxies outermost regions seem to merge as the Clouds are at their current natural separation. This occurs in both $xy$ and $xz$ Cartesian planes shown in Fig.~\ref{fig:rrl-sep}. This simulation shows that the effect of merging contours is natural for galaxies that are close enough. It does not necessarily imply that there is an additional structure between these objects, i.e. the bridge, as the model itself has proven that there is no overdensity located in the genuine Bridge area.

However, one can argue about the lowest contours being more spread in the direction toward the Magellanic Bridge than in any other direction (in every plot in Fig.~\ref{fig:rrl-sep}). In order to verify this we would need to significantly improve our analysis and this is beyond the scope of this paper. Firstly, we would need to abandon the $\sigma$-clipping and choose another method of rejecting outliers that would take into account the real error distribution which is not normal in the case of PL relations \citep{Nikolaev2004,Deb2018}. Using $\sigma$-clipping we probably remove some of the objects that are truly located at lower and higher distances in the outskirts of the LMC and SMC. Thus, the lowest contours perpendicular to the line of sight should not be used in such detailed analysis. Secondly, we would need to observe the entire LMC outskirts located in the eastern, northern and southern directions. Even though the OGLE has lately significantly improved its sky coverage in the Magellanic System, it is still not sufficient for such analysis where we need to compare the very lowest contours.

Summarizing this subsection, we want to emphasize that comparison of the lowest level contours is not sufficient to state whether there exists a bridge-like connection between the Magellanic Clouds or not.

% % % % % %
\section{A Reanalysis}

The results that we presented in the previous section are in perfect agreement with our findings from \citealt{PaperII}. We do not see any evident connection in the Magellanic Bridge area but only two extended structures in the LMC and SMC outskirts that are overlapping. Lately, \citealt{B17} have also presented a map of the OGLE RRL stars in the Magellanic System (their Fig.~18). This map clearly shows a connection between the Magellanic Clouds that was supposed to be consistent with {\it Gaia} DR1 RRL candidate distribution presented in their paper. This seems to be in contradiction with any of our results -- for comparison see Fig.~16 from \citealt{PaperII} or Fig.~\ref{fig:rrl-cart} in this paper. We tried to reconstruct results from \citealt{B17}. In this subsection we describe the method that we used to reanalyse the OGLE sample of RRab stars.

\subsection{The Influence of Coordinate System}

First, we have transformed our data to the Magellanic Bridge coordinate system which was used by \citealt{B17} and which is described in their paper (see Section~2.2 therein). This is simply a rotated equatorial coordinate system with a northern pole at $\alpha_{\rm p}=2^{\rm h}38^{\rm m}00^{\rm s}$, $\delta_{\rm p}=15\arcdeg28\arcmin30\arcsec$. In this coordinate system both LMC and SMC centers are located on the equator, thus it aligns well with the Magellanic Bridge \textsc{H i} structure (\citealt{B17}). The LMC center is located at $(X_{MB},Y_{MB})=(0,0)$.

\begin{figure}[htb]
	\begin{center}
	\includegraphics[width=.47\textwidth]{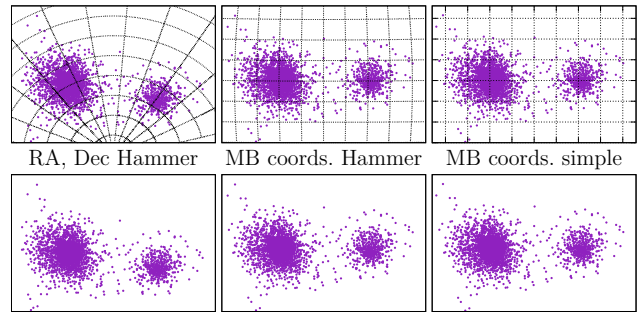}
	\end{center}
	\caption{The plot shows $\sim9\%$ of the OGLE RRL entire sample in different coordinate systems with different sphere projections. Bottom panels show the same sample and projection as top but without a grid.}
	\label{fig:rrl-coord}
\end{figure}

To test differences between each coordinate system and each projection used we plotted our entire sample of RRab stars in six different ways in Fig.~\ref{fig:rrl-coord}. Left column shows equatorial coordinates and Hammer equal-area projection, middle -- Magellanic Bridge coordinates with a Hammer equal-area projection, and left -- also Bridge coordinates but with $X_{MB}$ plotted on the $x$ axis, and $Y_{MB}$ on the $y$ axis. Bottom panel shows the same data as the top but without a grid.

It may seem from the top panel that the Magellanic Clouds are differently rotated toward each other in the first two plots, which are comparing the equatorial and Bridge coordinates. The bottom panel reveals that it is only an optical illusion created by the coordinate grid. Indeed, the entire sample is only slightly rotated but no compression occurs.

We also compare two different sphere projections of the Magellanic Bridge coordinates: Hammer-equal area shown in middle column with a projection used by \citealt{B17} shown in right column. Note that the latter is not equal-area and this influences the plot. The effect is not significant though, especially in the Bridge area which is located very close to the equator.

\subsection{No Evident Connection}

In order to thoroughly check whether we actually see the connection in the OGLE data we have reanalyzed the entire sample of RRab stars. To reproduce \citealt{B17} map from their Fig.~18 as precise as possible, we have once more calculated metallicities and distance moduli using the same technique as they have (V. Belokurov, private communication). In the next paragraphs we describe this method and later we discuss our results.

We used \citet{Smolec2005} relation for OGLE \textit{I}-band to calculate metallicity of each RRL star. This relation was derived for Fourier sine decomposition and \citet{Soszynski2016} gives coefficients for the cosine decomposition, thus we transformed the $\phi_{31}$ coefficient before applying \citet{Smolec2005} relations:
\begin{equation}
	\phi_{31,\sin}=\phi_{31,\cos}+\pi
\end{equation}
And the relation is (Eq.~2 from \citealt{Smolec2005}):
\begin{equation}
	[{\rm Fe}/{\rm H}]=-3.142-4.902P+0.824\phi_{31}
\end{equation}

Then we transformed $[{\rm Fe}/{\rm H}]$ to $Z$ using Eqs.~9 and 10 from \citet{Catelan2004}:
\begin{equation}
	\log Z=[{\rm Fe}/{\rm H}]+\log (0.638f+0.362)-1.765
\end{equation}
where $f=10^{\left[\alpha/{\rm Fe}\right]}$. We assumed $[\alpha/{\rm Fe}]=0$ following \citealt{B17}, although \citet{Carney1996} suggested $[\alpha/{\rm Fe}]=0.30$ based on stellar clusters. We have tested both options in our analysis and found that this value does not influence our main conclusions. Then we used theoretical calibrations of the PL relations from \citet{Catelan2004} to calculate absolute magnitudes of the RRab stars. Their Eq.~8 shows quadratic dependency between metallicity and absolute \textit{V}-band magnitude:
\begin{equation}
	M_V=2.288+0.8824\log Z+0.1079(\log Z)^2
\end{equation}
And Eq.~3 from \citet{Catelan2004} for the $I$-band absolute magnitude:
\begin{equation}
	M_I=0.4711-1.1318\log P+0.2053\log Z,
\end{equation}
where $P$ is the fundamental mode pulsation period.

Having absolute magnitudes we were able to calculate color excesses:
\begin{equation}
	E(V-I)=m_V-m_I-(M_V-M_I)
\end{equation}
where $m_{V,I}$ are observed mean magnitudes. We used value obtained by \citet{Nataf2013}, ${\rm d}A_I/{\rm d}(E(V-I))=1.215$, and assumed that $A_I=1.215E(V-I)$. Note that these values were obtained for the Galactic bulge where the extinction is nonuniform and anomalous (standard extinction is around 1.5, see \citealt{Udalski2003}). However, we decided to apply values from \citet{Nataf2013} in order to exactly follow the procedure used by \citealt{B17}.

In the last step we calculated distance moduli using magnitudes in the \textit{I}-passband:
\begin{equation}
	\mu=m_I-M_I-A_I
\end{equation}

\begin{figure}[htb]
	\includegraphics[width=.42\textwidth]{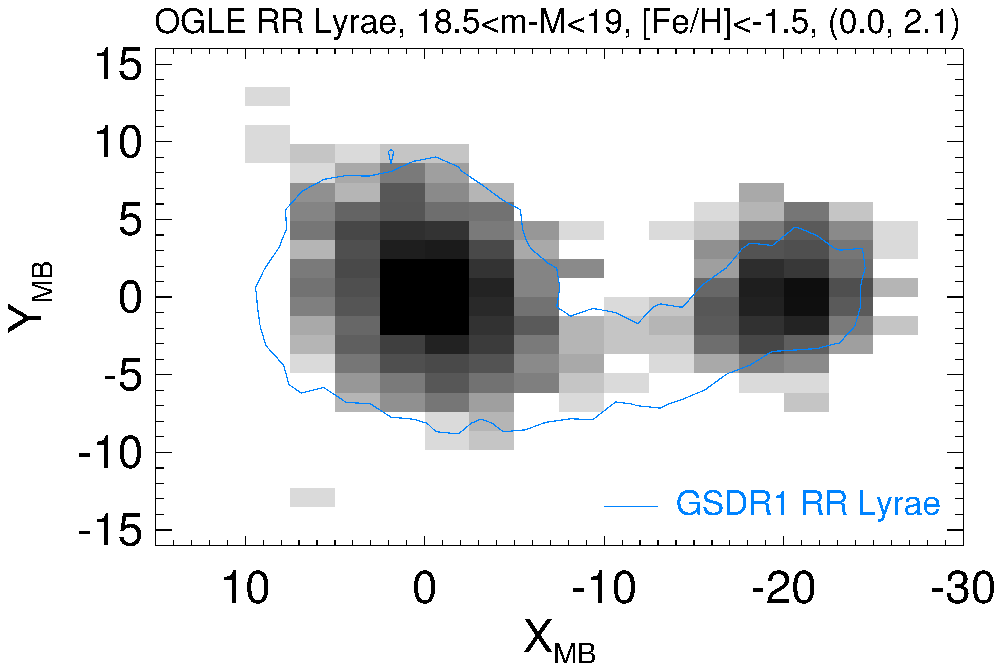} \\
	\includegraphics[width=.47\textwidth]{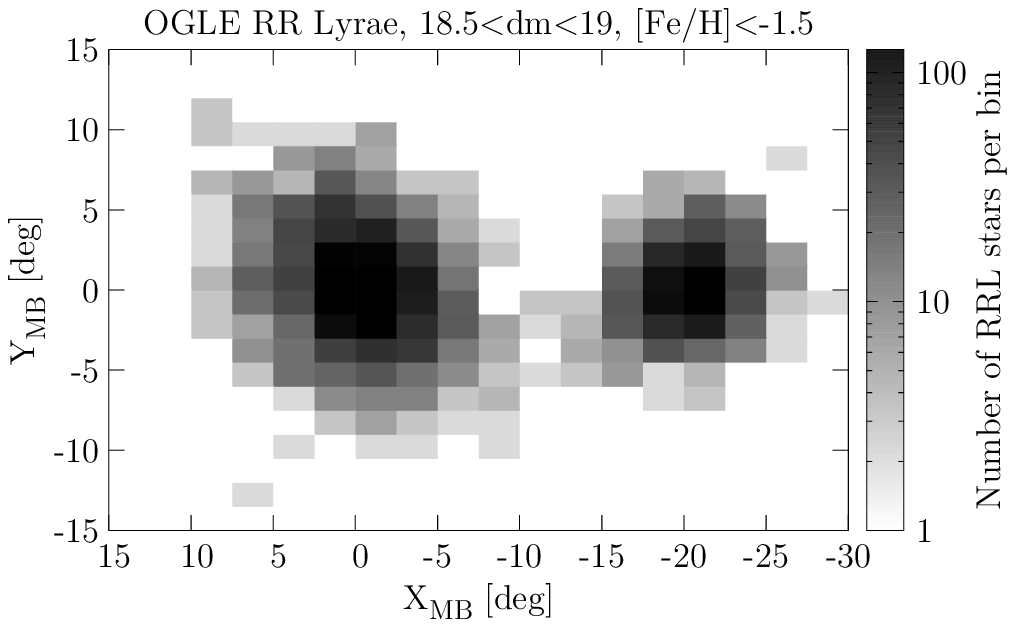} \\
	\caption{{\it Top:} Bottom panel of Fig.~18 from \citealt{B17} showing OGLE RRab stars in the Magellanic System. The data are binned into rectangles and gray color-scale is applied to show the column density. Only shown are RRab stars with distance moduli falling into range $18.5<m_I-M_I<19$ and metallicities $[{\rm Fe}/{\rm H}]<-1.5$. The scale is logarithmic and is limited from $10^0$ to $10^{2.1}$ RRab stars per square degree. Blue contour represents the density of {\it Gaia} DR1 RRL candidates analysed by \citealt{B17}. The coordinates used are in the Magellanic Bridge system and the sky projection is not equal-plane. {\it Bottom:} Our map showing OGLE RRab stars in the Magellanic System with parameters calculated the same method as in \citealt{B17}. Note that the bridge-like structure is even better visible due to the elongation of bins along the connection (and along the equator).}
	\label{fig:rrl-beloog}
\end{figure}

The reproduced map is shown in the bottom panel of Fig.~\ref{fig:rrl-beloog} together with the original map from \citealt{B17} shown in the top panel. Both plots show the OGLE RRab sample though in the case of our map (bottom panel) we used the updated sample. Both plots present samples with the same cuts: distance moduli falling into range $18.5<m_I-M_I<19$ and metallicities $[{\rm Fe}/{\rm H}]<-1.5$, as well as other parameters: coordinates, sphere projections, method of calculation, bin sizes and ranges, color-scale range. Under all of these conditions we were able to reproduce the connection visible in \citealt{B17} map. The bridge-like structure is visible only on a very low level of counts. Moreover, due to large bin size and elongation of bins along the $y$ axis, thus along the Bridge (and also along the equator), the connection is even more pronounced.

\begin{figure*}[htb]
	\centering
	\includegraphics[width=.63\textwidth]{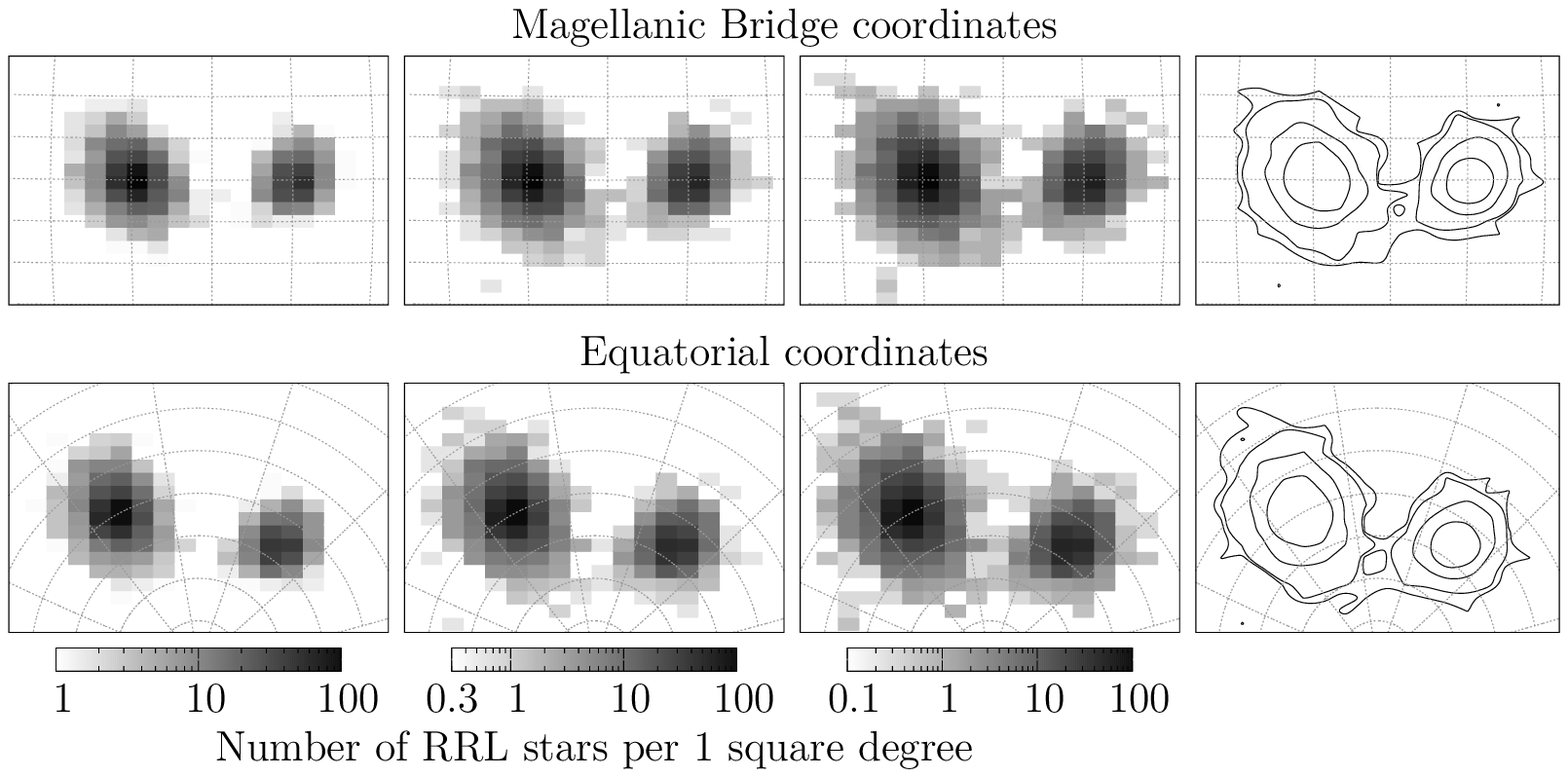} \\
	\includegraphics[width=.63\textwidth]{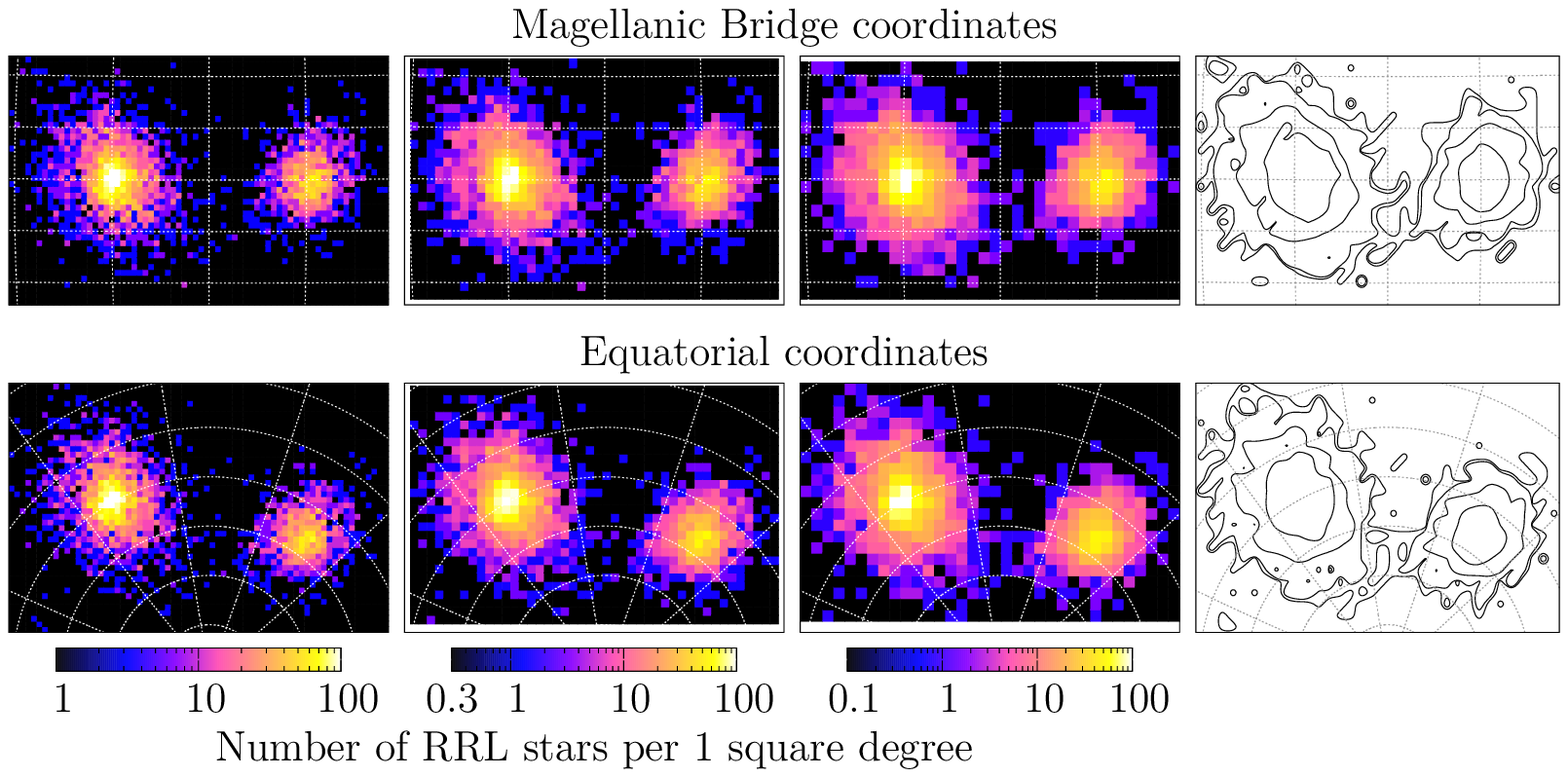} \\
	\caption{Every plot shows the same sample as in the bottom panel of Fig.~\ref{fig:rrl-beloog}. {\it Top:} Here we used the same binning as in Fig.~\ref{fig:rrl-beloog} but in Hammer equal-area projection applied to the Magellanic Bridge (top row) and equatorial (bottom row) coordinate system. Each column shows different bottom range of color-scale. The right column shows contours that are on the levels of 0.5, 1, 5, 15 RRab stars per 1 square degree. {\it Bottom:} We used square bins instead of rectangular ones. We also applied different colour-scale with different range to show the subtlest features. The bin size is linearly differing between each column. The top row shows Magellanic Bridge while the bottom -- equatorial coordinate system. Both are represented using Hammer equal-area projection. Additionally, the right column shows contours fitted to binning shown in the second column. The contours levels are: 1, 5, 15 RRab stars per 1 square degree.}
	\label{fig:rrl-bin}
\end{figure*}

To test whether the choice of coordinates system also influences the visibility of the bridge-like connection we also plotted the same sample as in Fig.~\ref{fig:rrl-beloog} using different transformations. Top panels of Fig.~\ref{fig:rrl-bin} show the same rectangular bins with a grey color-scale but using an equal-area Hammer projection applied to the Magellanic Bridge (top row) and equatorial (bottom row) coordinate systems.

In the left column of Fig.~\ref{fig:rrl-bin}, where the color-scale range starts at 1 star per 1 square degree, the connection is not visible in either coordinate system. It only starts to emerge in the second column where the bottom of color-scale range is under the level of 1 star per 1 square degree -- precisely at 0.3. The bridge-like structure is even more visible in the third column where the range is even lower. Although, in this plot also other extended features are starting to emerge. Moreover, comparing top and bottom grey rows leads to a conclusion that the connection is more clearly visible in the Magellanic Bridge coordinates. This is due to the fact that in this system the bridge-like structure is located along the equator. Comparing contours for both coordinate systems we conclude that the contours do connect in both cases but on a very low level. Again, the connection is slightly more visible in the Magellanic Bridge coordinate system.

Furthermore, to test whether the binning influences the results, we also plotted the same sample using square bins of different sizes. Results are shown in bottom panels of Fig.~\ref{fig:rrl-bin}. Similarly to the grey panels, the top row shows Magellanic Bridge and bottom -- equatorial coordinates. Comparison of rectangular and square bins leads to a conclusion that binning indeed has an impact on the visibility of the bridge-like structure. The square bins make the connection appear significantly less visible than rectangular bins. It is not a surprise, as the rectangular bins used by \citealt{B17} were aligned with the bridge.

% % % % % %

\section{\citealt{B17} RRL Candidates from {\it Gaia} DR1}

% %
\subsection{Selection Process}

In this Section we present results of an analysis of the {\it Gaia} DR1 data \citep{Gaia2016} performed the same way as in \citealt{B17}. The main goal of \citealt{B17} was to select RRL candidates from {\it Gaia} DR1 and analyze the on-sky distribution of these stars in the Magellanic System area, with an emphasis on the Bridge. They found that there exist an evident connection between the Magellanic Clouds. Hereafter we try to reproduce their results and compare with OGLE and {\it Gaia} DR2 database.

In order to reproduce \citealt{B17} list of RRL candidates using {\it Gaia} DR1, we use their procedure with the following steps: 
\begin{enumerate}
	\item From the entire {\it Gaia} DR1 database we selected all sources located in an area where RA$\in (0^{\rm h},9^{\rm h})\cup(22^{\rm h},24^{\rm h})$ and Dec $\in (-85\arcdeg,-45\arcdeg)$ with more than 70 CCD crossings and Galactic longitude $b \leq -15\arcdeg$. The latter two requirements are corresponding to {\it iv} and {\it vii} cuts from \citealt{B17} (see their Section~3.3).
	\item An appropriate value of extinction $E(B-V)$ was found for all sources using \citet{Schlegel1998} maps. This allowed us to deredden all of the objects from the selected sample using following relation for extinction coefficient for {\it Gaia} $G$-band (Eq.~1 from \citealt{B17}), $A_G$:
	\begin{equation}
		A_G = 2.55 E(B-V).
	\end{equation}
	\item Then we calculated amplitude value, Amp, using the following relation (Eq.~2 from \citealt{B17}):
	\begin{equation}
		{\rm Amp} = \log_{10}\left(\sqrt{N_{obs}}\frac{\sigma_{\overline{I_G}}}{\overline{I_G}}\right),
	\end{equation}
	where $N_{obs}$ is the number of CCD crossings, $\overline{I_G}$ is the mean flux in {\it Gaia} $G$-band and ${\sigma_{\overline{I_G}}}$ is the error of the mean flux.
	\item Finally, the remaining cuts presented in Sec.~3.3 of \citealt{B17} were applied. The cuts concern: amplitude as defined above, Astrometric Excess Noise (AEN), $G$-band magnitude, reddening.
\end{enumerate}

We applied different versions of cuts {\it ii} and {\it vi} as presented in \citealt{B17}. We use both strict and weak cuts on the amplitude, $-0.75 < {\rm Amp} < -0.3$ and $-0.65 < {\rm Amp} < -0.3$ respectively. Similarly for the AEN, where $\log_{10}({\rm AEN})<-0.2$ is a strict cut, and $\log_{10}({\rm AEN})<-0.2$ is weak. Additionally, we also analyzed even weaker version of the AEN cut, where $\log_{10}({\rm AEN})<0.3$. Results are presented in Fig.~\ref{fig:rrl-dr1}.

% %
\subsection{Two-Dimensional Analysis}

\begin{figure*}[htb]
\centering
	\includegraphics[width=.9\textwidth]{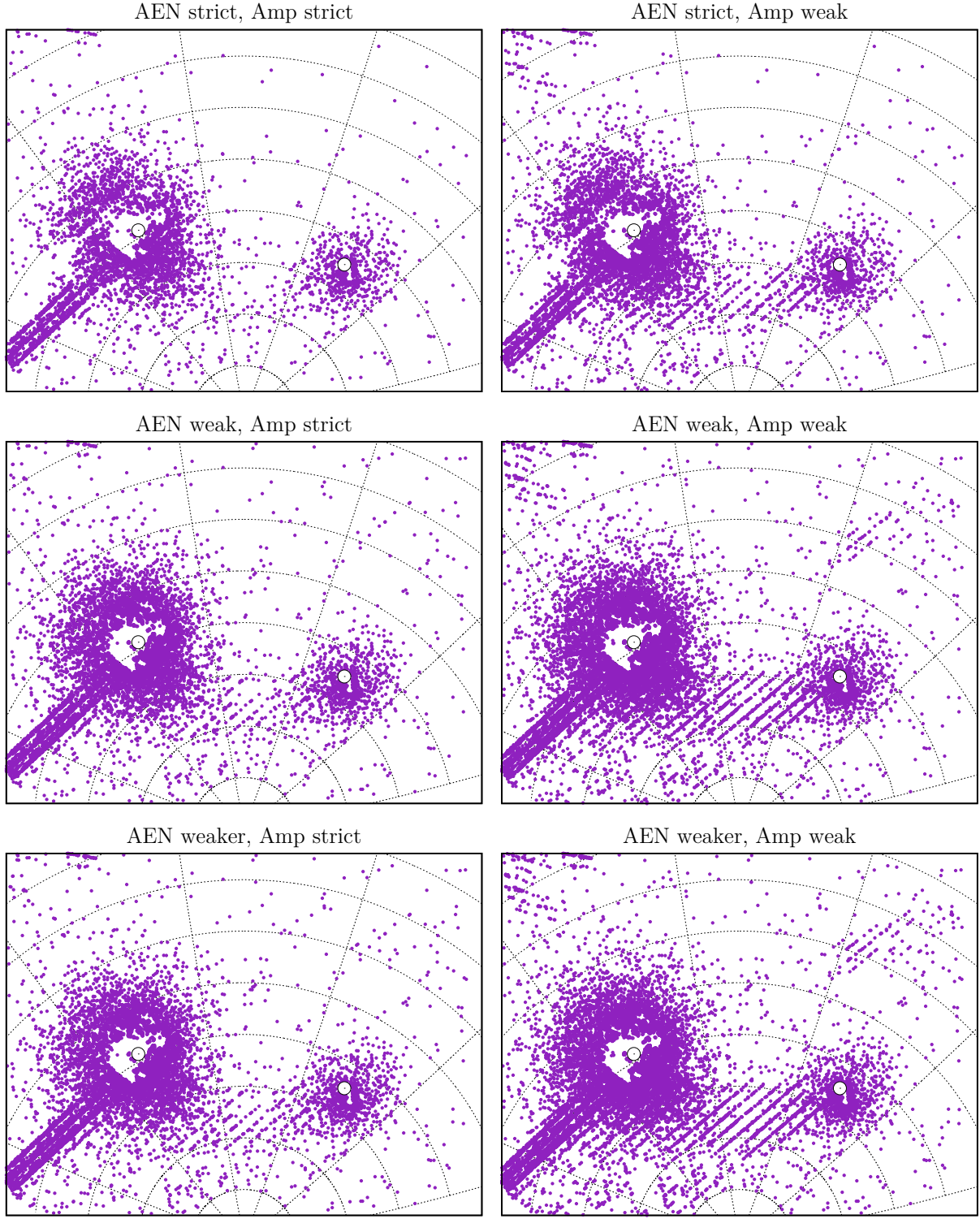}\\
	\caption{The on-sky locations of RRL candidates using different versions of \citealt{B17} cuts. Clearly visible is the non-physical artifact east of the LMC that we did not remove. It is created by spurious variables, which are caused by {\it Gaia} DR1 cross-match failures (\citealt{B17}). Stripes are matching {\it Gaia} scanning pattern. Similar stripes visible in the Bridge area suggest that many of objects located there are non-physical sources. Additionally, white circles mark the LMC \citep{vanderMarel2014} and SMC \citep{Stanimirovic2004} centers.}
	\label{fig:rrl-dr1}
\end{figure*}

Fig.~\ref{fig:rrl-dr1} clearly shows that when using the Amp and AEN cuts both in strict version there are not many stars left in between the Magellanic Clouds. To test whether this result reproduces the RRL bridge reported by \citealt{B17} we binned the data in the same way as their Fig.~11. The bins are on too low level and no connection is visible, thus strict cuts do not reproduce their bridge. Moreover, the sample we obtained consisted of $\sim 7\ 000$ objects which is three times less numerous than \citealt{B17} sample ($\sim 21\ 500$ objects). In case of applying at least one cut in weak version we obtained a distribution revealing stripes in the Bridge area.

Right panel of Fig.~5 in \citealt{B17} shows an on-sky distribution of all nominally variable stars selected from DR1. Many non-physical features are visible, including the artifact east of the LMC. \citealt{B17} perform a detailed analysis of stripes appearing in this plot (for details see their Sec.~3.2 and Fig.~6). These stripes are aligned with the {\it Gaia} scanning pattern and are caused by cross-match failures. Thus, most of sources forming the stripes are not physical. \citealt{B17} further claim that the stripes disappear due to the cuts applied, and only a small number of spurious sources fall into selected RRL regions. Our study reveals that this is not the case and the final RRL candidate sample still contains a number of non-physical sources, forming the stripes. Comparing our Fig.~\ref{fig:rrl-dr1} with Fig.~5 from \citealt{B17}, it is clearly visible that the features in the Magellanic Bridge area are not removed by the applied procedure. Thus, the discovery of the bridge-like connection by \citealt{B17} was based on a non-physical structure.

Moreover, in Fig.~\ref{fig:rrl-dr1} clearly visible is a non-physical artifact located east of the LMC that we did not remove. This feature is located in the area most influenced by cross-match failures in the {\it Gaia} DR1 (see masked pixels in the left panel of Fig.~5 in \citealt{B17}). The sources in between the Magellanic Clouds are forming stripes that are aligned with the non-physical artifact east of the LMC. This supports our conclusion from the previous paragraph that the Bridge area is highly influenced by non-physical sources. Additionally, we obtain an area close to the center of the LMC, where the sources are missing, due to the requirement of $N_{obs} > 70$. However, we managed to recreate the sample in the Magellanic Bridge which is our main area of interest.

\begin{figure}[htb]
	\includegraphics[width=.42\textwidth]{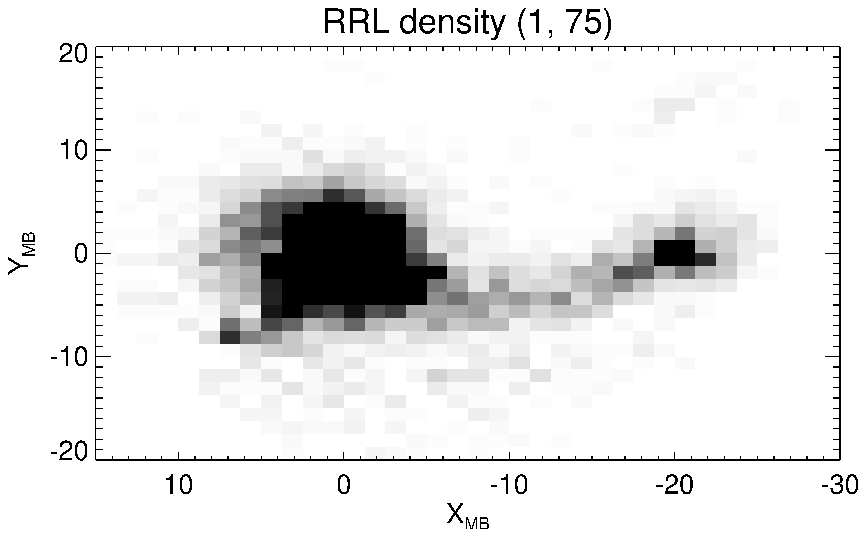} \\
	\includegraphics[width=.47\textwidth]{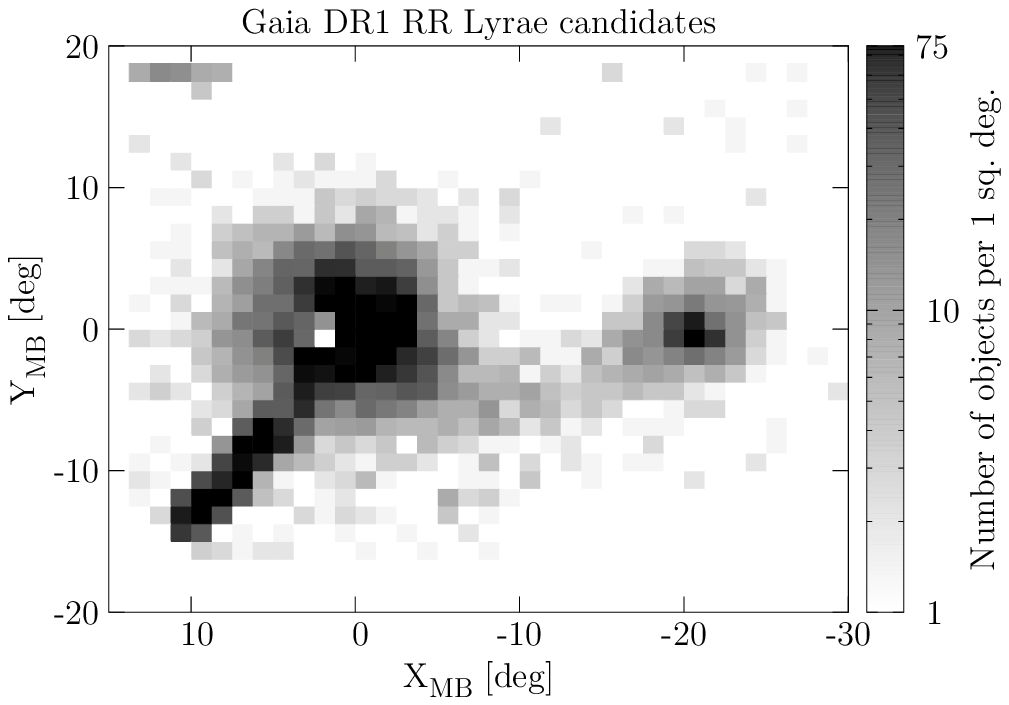} \\
	\caption{A comparison of top panel of Fig.~11 from \citealt{B17} (top panel) with the map obtained using the same technique (bottom panel). Bottom panel shows a sample with strict Amp cut and weaker AEN cut.}
	\label{fig:rrl-dr1belo}
\end{figure}

As our final sample of RRL candidates we selected the one with strict cut on Amp and weaker cut on AEN as it perfectly reproduced a sample of 113 central Bridge objects from \citealt{B17} analysis (V. Belokurov, private communication). In Fig.~\ref{fig:rrl-dr1belo} we show a comparison of a binned map of this sample with Fig.~11 from \citealt{B17}. Both maps are plotted using the same coordinate system, sphere projection, bin size, and color-scale range. We managed to reproduce Bridge features very well. One main difference between our map and \citealt{B17} is the non-physical artifact located east of the LMC. Note that in this binning the {\it Gaia} stripes are not visible. Our sample contains more than 13 300 stars. This is more than half of \citealt{B17} sample, indicating that they have applied even weaker cuts in their final sample.

\begin{figure*}[htb]
	\includegraphics[width=\textwidth]{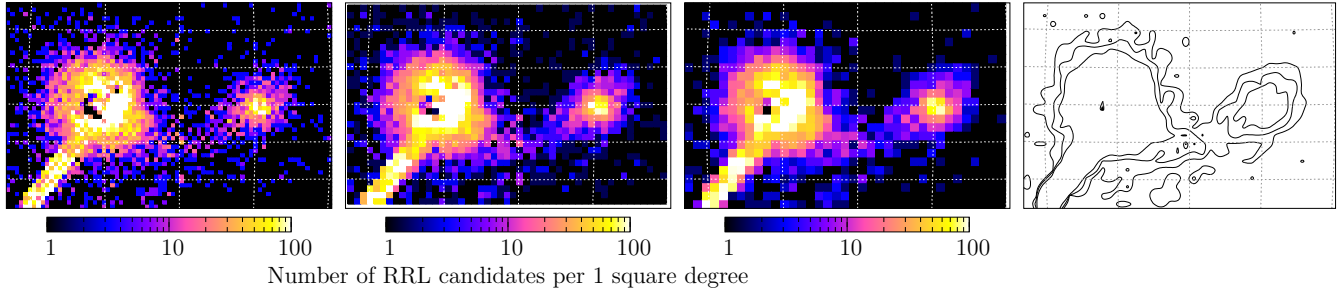}
	\caption{The same sample as in Fig.~\ref{fig:rrl-dr1belo} bottom panel but binned using square bins.}
	\label{fig:rrl-dr1bin}
\end{figure*}

In Fig.~\ref{fig:rrl-dr1bin} we also show our final sample using square bins of different sizes. We represented the data in the Magellanic Bridge coordinates using Hammer equal-area projection. As the bin size increases from left to right the {\it Gaia} stripes appear less visible. The contours shown in the right panel match very well contours obtained by \citealt{B17} (see their Fig.~12).

% %
\subsection{Comparison with OGLE and {\it Gaia} DR2}

The OGLE Collection of RRL Stars in the Magellanic Clouds is nearly complete -- the level of completeness is higher than 95\% \citep{Soszynski2016,Soszynski2017}. Therefore, we cross-matched the list of RRL candidates obtained in this section with the OCVS to test how many of these objects are genuine RRL stars. We cross-matched separately the entire sample of \citealt{B17} DR1 RRL candidates and a subsample created by selecting only objects in the Bridge area located between the LMC and SMC centers. This Bridge subsample consists of sources located within $-20\arcdeg<X_{MB}<0\arcdeg$.

Results are presented in Tab.~\ref{tab:rrl-dr1xm} which shows that only about $41.4\%$ of objects in the entire RRL candidates sample are genuine RRL stars. For the Bridge subsample this ratio is at the level of about $47.5\%$. Moreover, we separately tested a subsample of 113 objects in the central Bridge area, where the \citealt{B17} overdensity is located. Only 17 among these objects are RRL stars, which leads to a total ratio of $15.0\%$. The difference between this ratio for the entire sample and central Bridge subsample indicates a higher contamination in the latter. This is consistent with the fact that many sources in the Bridge area are non-physical. The contamination of 85\% in the central Bridge sample is not consistent with \citealt{B17}, who give a value of $30-40\%$ for their entire sample.

\begin{table}[htb]
\caption{\citealt{B17} RRL candidates from {\it Gaia} DR1 -- cross-match}
\label{tab:rrl-dr1xm}
\centering
\begin{tabular}{lrDDDD}

\multirow{2}{*}{Sample} & \multirow{2}{*}{No. obj.} & \multicolumn{8}{c}{Cross-match with} \\
                        &  						    & \multicolumn{4}{c}{OGLE RRL}  & \multicolumn{4}{c}{{\it Gaia} DR2 RRL} \\
%\hline
\cline{1-10}
Entire   & 13327 & 5516 & (41.4\%) & 4872 & (36.6\%) \\  % 47685 in OCVS
MBR      &  6041 & 2971 & (47.5\%) & 2542 & (42.1\%) \\  % 23996 in OCVS
Cen. MBR &   113 &   17 & (15.0\%) &   15 & (13.3\%) \\
%\hline
\cline{1-10}
\multicolumn{10}{p{.48\textwidth}}{The MBR sample constitutes of objects located between $-20\arcdeg<X_{MB}<0\arcdeg$. The central MBR sample are objects located between the Magellanic Clouds that contribute to \citealt{B17} overdensity.}
\end{tabular}
\end{table}

Note that the area that we use for the RRL candidates selection process is larger than OGLE-IV fields coverage (see Fig.~1 in \citealt{PaperIII}). For the entire sample the difference in purity level is larger than for the Bridge sample, as the former includes also the non-physical artifact in DR1 data that is not entirely covered by the OGLE fields. For the Bridge sample only a few sources are located north and south of the OGLE footprint. Thus, this effect should not be significant for the selected Bridge subsample. It also explains the significant difference between the cross-matches of RRL candidates samples with the OGLE data.

We would expect that a proper technique of selecting RRL candidates would lead to a result of high completeness. To test that, we compared the number of RRL stars from our reconstructed sample using the described technique to the total number of these objects in the OGLE database in the Magellanic System. The entire RRL candidates list has a completeness level of 11.6\%, while for the Bridge sample -- 12.4\% which is consistent with what \citealt{B17} estimated. This means that almost 90\% of RRL stars located in the OGLE-IV fields in the Magellanic System were not discovered.

\begin{figure}[htb]
	\includegraphics[width=.47\textwidth]{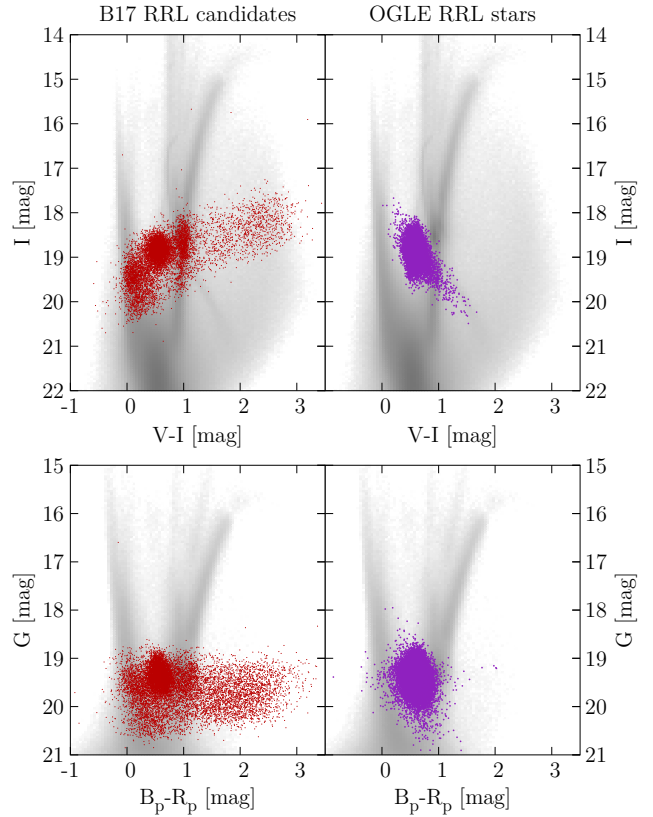}
	\caption{CMDs of \citealt{B17} RRL candidates (left, red) obtained in this section and the cleaned sample of OCVS RRL stars (right, purple) overplotted on the Hess diagrams for the data from selected fields in the Magellanic System. \textit{Top:} OGLE photometry. \textit{Bottom:} {\it Gaia} DR2 photometry.}
	\label{fig:rrl-cmdbelo}
\end{figure}

Moreover, we also cross-matched the obtained RRL candidates lists with the entire OCVS published up to date and with the entire OGLE database. About 2.3\% of objects from the candidate samples are eclipsing binaries. A few are also classified in the OCVS as long period variables. We show in Fig.~\ref{fig:rrl-cmdbelo} a comparison of color-magnitude diagram (CMD) of the sample obtained in this section with the cleaned sample of RRL stars. Both are overplotted on the OGLE data (top panel) and {\it Gaia} DR2 data (bottom panel) from selected fields in the Magellanic System. The reconstructed \citealt{B17} sample spans toward different areas than those usually occupied by the genuine RRL stars. Thus, this sample contains a lot of different types of objects.

We have also performed a cross-match between the RRL candidate sample from {\it Gaia} DR1 obtained in this section with the {\it Gaia} DR2 RRL stars listed in \texttt{vari\_rrlyrae} table \citep{Gaia2018,Holl2018,Clementini2019}. Tab.~\ref{tab:rrl-dr1xm} lists the exact results. Only about $37\%$ of sources from the RRL candidates sample are present in the {\it Gaia} DR2. For the Bridge sample this result is slightly higher -- $42\%$. Lower numbers as compared to the cross-match with the OGLE data are probably a result of lower DR2 RRL sample completeness that we describe in the following section.

% % % % % %

\section{{\it Gaia} DR2 RRL Stars in the Bridge}

% %
\subsection{Comparison with OCVS}

\begin{figure}[htb]
	\begin{center}
	\includegraphics[width=.47\textwidth]{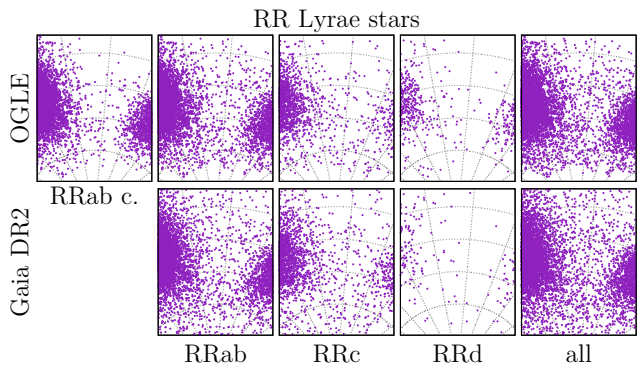}
	\end{center}
	\caption{Comparison of OGLE (top row) and {\it Gaia} DR2 (bottom row) RRL stars in the Magellanic Bridge area.}
	\label{fig:rrl-dr2}
\end{figure}

In \citealt{PaperIII} we already presented a comparison of {\it Gaia} DR2 Cepheids with OGLE Collection of Cepheids. In this paper we show a similar discussion concerning RRL stars. We again focus on the Magellanic Bridge area. Fig.~\ref{fig:rrl-dr2} shows a comparison of on-sky distributions of different types of RRL stars. The OCVS data is presented in the top row, while DR2 -- bottom. First column shows the cleaned RRab sample that we calculated only for the OCVS. Other columns compare RRab entire samples, RRc and RRd. In the last column we plotted all types of RRL stars together. The difference is visible only for RRd stars where OCVS seems to contain more objects than DR2 in the Bridge area. Other OCVS plots show the lack of objects in the southern parts which is caused by the OGLE sky coverage truncation. Moreover, different areas not covered by the OGLE footprint are visible, i.e. at ${\rm RA}\approx65\arcdeg$.

Using our updated sample of RRL stars and the {\it Gaia} DR2 sample we performed a cross-match between these two. Similarly to \citealt{PaperIII}, we selected a DR2 sample covering the entire OGLE fields in the Magellanic System. In this area {\it Gaia} DR2 has a completeness of 69.0\% for all RRL stars. This value is consistent with Tab.~2 in \citet{Holl2018}. Again, this is not surprising as the OGLE Collection of RRL stars was a training set for the {\it Gaia} selection algorithms.

We also cross-matched the {\it Gaia} DR2 RRL stars with objects of this type that we identified in the newly added OGLE fields located east and north of the LMC. This data are not yet published, thus we tested how effective DR2 classification is. The completeness for this subsample is of 70\% and is virtually the same as for the entire sample.

% % % % % %

\section{Comparison of Different Tracers Distribution}

\begin{figure*}[htb]
	\includegraphics[width=\textwidth]{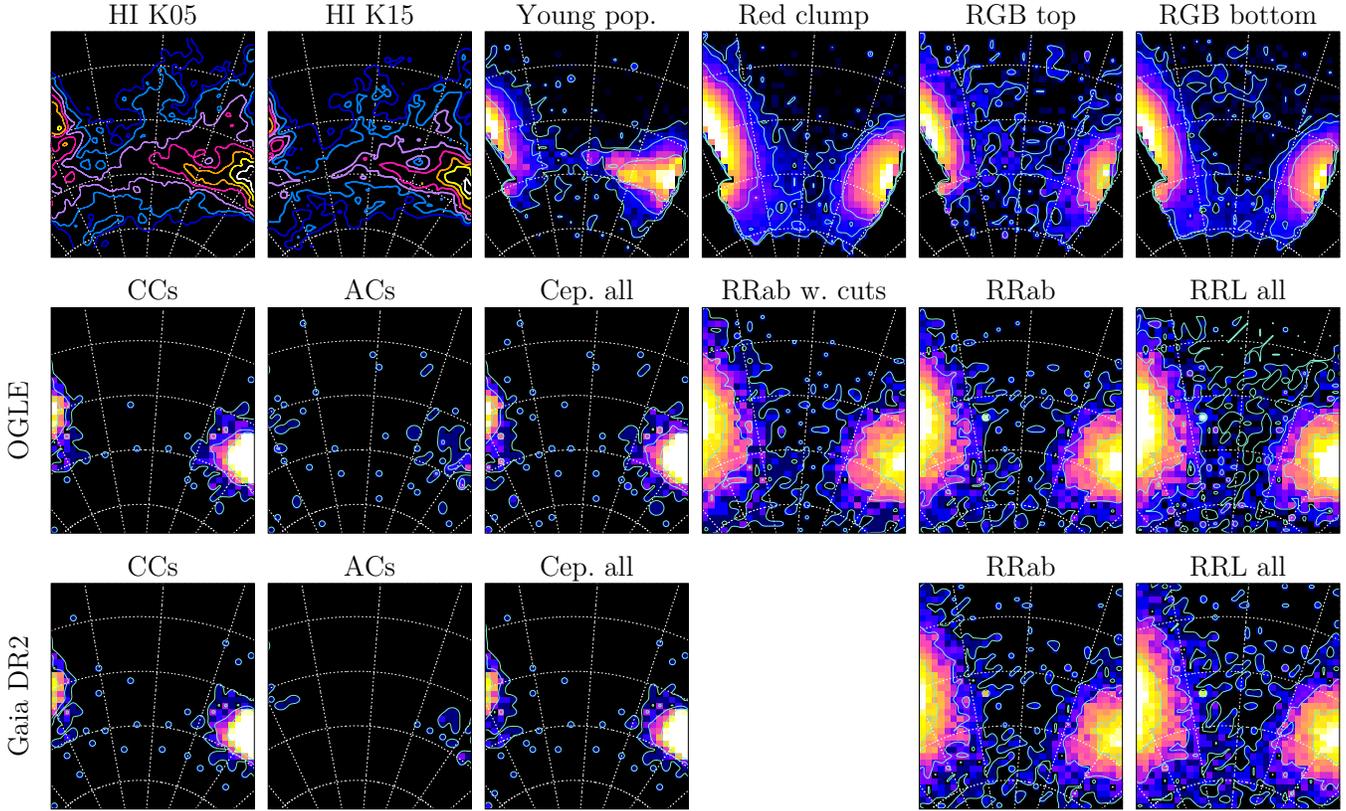}
	\caption{Comparison of on-sky locations of different tracers in a Hammer equal-area projection. Each plot has its own color scale and contours levels. {\it Top row:} First panel on the left shows neutral hydrogen density contours from Leiden/Argentine/Bonn \textsc{H i} Survey (\citealt{Kalberla2005}, the same as in Fig.~8 in \citealt{Skowron2014}, see this Fig. description for details). Second panel shows \textsc{H i} from Galactic All Sky \textsc{H i} Survey \citep{McClure-Griffiths2009,Kalberla2010,Kalberla2015}. Contours are on the levels $(1,2,4,8,16,32,64)\cdot10^{20}\ {\rm cm^{-2}}$. On both panels the \textsc{H i} is integrated over velocity range $80<v<400\ {\rm km\ s^{-1}}$. Panels from third to fifth show column densities of different stellar populations as selected in the color-magnitude diagrams in \citet{Skowron2014}. Shown here for comparison are young population, red clump objects, top and bottom of red giant branch (RGB). {\it Middle row:} Classical pulsators from the OCVS. {\it Bottom row:} Classical pulsators from the {\it Gaia} DR2.}
	\label{fig:comp}
\end{figure*}

In this section we compare on-sky distributions of different tracers in the Magellanic Bridge area. The main plot that we discuss is shown in Fig.~\ref{fig:comp}. First row contains the \textsc{H i} density contours from Leiden/Argentine/Bonn \textsc{H i} Survey (\citealt{Kalberla2005}, the same as Fig.~8 in \citealt{Skowron2014}) and from Galactic All Sky \textsc{H i} Survey \citep{McClure-Griffiths2009,Kalberla2010,Kalberla2015}, young population, red clump, top and bottom of red giant branch (RGB) distributions (Figs.~8, 9, 11, 13 from \citealt{Skowron2014}). Middle row shows different types of classical pulsators from the OCVS that we investigated in \citealt{PaperIII} and this paper, namely classical Cepheids (CCs), anomalous Cepheids (ACs), both these types plotted together, RRab the cleaned sample, RRab the entire sample, RRL all types plotted together. Similarly, these types of objects are shown in the bottom row using data from {\it Gaia} DR2 (with the exception of the cleaned RRab sample that we calculated only for the OCVS). All of these plots show a color-coded column density, while lines represent density contours. For each plot the color scale and contours levels are different.

Comparing neutral hydrogen with other maps it is clearly visible that the most matching are distributions of young stars and CCs. Each of these three seems to follow a bridge-like connection between the Magellanic Clouds along similar declination range: ${\rm Dec}\in (70\arcdeg,72\arcdeg)$. Older tracers are more spread and do not follow such strict connection. Red clump and RGB bottom stars are more concentrated in the southern parts of the Bridge than RGB top and RRL stars. RGB top objects are very spread and the lowest density contours show some clumps with the most populated stripe located along the young population bridge. The connection is though on a too low level to enable us to state that we see a connection similar to the young bridge. Summing up, for all intermediate-age and older tracers from \citet{Skowron2014} we can see two extended structures overlapping with no evident bridge-like connection.

RRL stars on-sky distribution shows that these stars are very spread in many directions -- even more than the other tracers that we discussed in the previous paragraph. Among the presented distributions, the distribution of RGB stars is the most similar to the distribution of RRL. The difference between RRab cleaned and entire samples shows that a number of objects is rejected from the Bridge sample. Note that, however, the column density in this area is low and removing even a small number of objects can result in significantly different density contours distribution. The entire RRab sample is distributed very similarly to the all RRL types, though the lowest density contours are slightly different. This is caused by the fact that the entire RRL sample is more numerous. Moreover, one can state that the ACs are similarly spread as the intermediate-age and older tracers. On the other hand, the ACs sample is significantly less numerous. We do not discuss further differences or similarities between different types of classical pulsators in this paper -- for a detailed statistical study of three-dimensional distributions see \citet{Iwanek2018}.

Fig.~\ref{fig:comp} shows that in DR2 many ACs were classified as CCs. This is the main reason for differences between the OCVS and DR2 Cepheids distributions. For a detailed description see Sec.~7 in \citealt{PaperIII}. Note also that \citet{Ripepi2018} lately reclassified DR2 sample of Cepheids. For a comparison see Fig.~12 in \citealt{PaperIII}. The {\it Gaia} DR2 RRL stars are distributed very similarly to the OGLE RRL -- both RRab and all types of these pulsators. These objects are very spread and while the lowest density contours do connect, it occurs on a very low level, below 1 star per 1 square degree. Thus, this cannot be the reason for stating that we see an evident bridge-like connection -- we actually do not.

% % % % % %

\section{Conclusions}

% dodatkowe gwiazdy w MBR - naturalna struktura, bo wchodzą gwiazdy z dwóch galaktyk, dwa nakładające się halo

In this paper, closely following our analysis of Cepheids in the Magellanic Bridge area \citep{PaperIII}, we present a detailed study of RR Lyrae stars in between the Magellanic Clouds using an extended OGLE Collection of Variable Stars (\citealt{Soszynski2016,Soszynski2017}, in prep.). We calculated absolute Wesenheit magnitudes for each RRL star, starting with estimating photometric metallicities \citep{Nemec2013}, and applying \citet{Braga2015} relations. This led us to calculating individual distances for our sample the same technique as we did in \citealt{PaperII} and \citet{Skowron2016}.

We analyzed three-dimensional distribution of RRL stars between the Magellanic Clouds in Cartesian coordinates. We show, confirming results from \citealt{PaperII}, as well as \citet{WagnerKaiser2017}, that we do not see an evident connection between the Magellanic Clouds in RRL stars. Objects located in the Bridge area form a smooth transition between the Clouds, rather than a bridge-like connection. The RRL distribution seems to represent two extended structures overlapping (i.e. halos or extended disks of the LMC and SMC). Additionally, we bin the data and show that the contours do connect, though on a very low level (below 1 star per 1 square degree or 1 kpc$^2$). It is too low to state that there exists an overdensity.

To test our sample numerically, we perform a multi-Gaussian fit. We made only two assumptions -- a number of Gaussians and a number of points to be simulated. Our results show that there is no Gaussian centered in the Bridge area. Thus, there is no additional population or overdensity therein. We also used the multi-Gaussian procedure to show, that when we separate the Magellanic Clouds by 8 kpc along the Cartesian $x$ axis, and then gradually shift the LMC and SMC back together, the lowest density contours start to connect at some point. Thus, the fact that the contours do connect, is not necessarily an evidence of an existence of an old bridge, as any contours will connect when the galaxies are close enough.

Moreover, to carefully study the lowest density contours, one needs to use a very precise technique to classify and analyze RRL stars. Even though the method we use is quite robust, as it is used in many different studies of three-dimensional structure, we do not think that it is precise enough to test the very outskirts of the Magellanic Clouds.

Lately, \citealt{B17} presented distribution of OGLE RRL stars in the Bridge that revealed a bridge-like connection (see their Fig.~18). This is in contradiction with results from \citealt{PaperII} or even from this paper that were described earlier. We reanalyzed our OGLE sample using different technique to test consistency. We show that the way the data is plotted influences the final impression. Carefully testing how the sample looks like in different coordinate systems and using different bin sizes, and types of bins, we show that we are able to reproduce \citealt{B17} plot only under specific conditions. Thus, because the connection is not always visible, we are even more convinced that it is on a very low level.

Using the same method as \citealt{B17} we also reproduced their main results by selecting RRL candidates from {\it Gaia} DR1 data. We applied a series of cuts to the data, as presented in \citealt{B17}. When all of the selection methods are used in strict versions, we obtain a very small number of objects in between the Magellanic Clouds. On the other hand, if at least one cut is weaker, the resulting distribution contains many spurious sources in the MBR area. Thus, we conclude that we are not able to reproduce \citealt{B17} RRL bridge without non-physical artifacts and we do not agree with their statement that the cuts presented remove most of the spurious sources. We also present a map of selected objects showing very evident stripes that, according to \citealt{B17}, match {\it Gaia} overlapping fields. This non-physical overdensity is matching very well the \citealt{B17} discovery. In the central Bridge area only 15\% of the sample are genuine RRL stars.

We also show, for the first time, the distribution of {\it Gaia} DR2 RRL stars in the MBR and compare it to the OCVS. On-sky locations of RRL stars from both samples are very consistent. Similarly to the OCVS RRL stars, the DR2 sample reveals a very spread distribution that rather resembles two overlapping structures than a strict bridge-like connection. The lowest density contours do connect, though on a very low level, again below 1 star per 1 square degree. These contours look slightly different when using only RRab stars than the entire RRL samples. This is probably due to the latter being more numerous. Again, we conclude that existence of a bridge-like structure should not be based on the lowest density contours.

At the same time we want to emphasise that we do not state that the RRL bridge does not exist. There are different surveys showing that there are some substructures in between the Magellanic Clouds. This is in agreement with our own study, as we also show that there are RRL stars in the Bridge area, though their distribution is rather not bridge-like and the overdensity is on a very low level.

% % % % % %

\acknowledgments

A.M.J.-D. is supported by the Polish Ministry of Science and Higher Education under ``Diamond Grant'' No. DI2013 014843 and by Sonderforschungsbereich SFB 881 "The Milky Way System" (subproject A3) of the German Research Foundation (DFG). P.M. acknowledges support from the Foundation for Polish Science (Program START). The OGLE project has received funding from the National Science Centre, Poland, grant MAESTRO 2014/14/A/ST9/00121 to A.U.

We would like to thank all of those, whose remarks and comments inspired us and helped to make this work more valuable. In particular we would like to thank Richard Anderson, Abhijit Saha, Vasily Belokurov, Anthony Brown, Laurent Eyer, Martin Groenewegen, Vincenzo Ripepi, Rados\l{}aw Smolec, Martino Romaniello, Krzysztof Stanek.

This research was supported by the Munich Institute for Astro- and Particle Physics (MIAPP) of the DFG cluster of excellence "Origin and Structure of the Universe", as it benefited from the MIAPP programme "The Extragalactic Distance Scale in the {\it Gaia} Era" as well as International Max Planck Research School (IMPRS) Summer School on "{\it Gaia} Data and Science 2018".

This work has made use of data from the European Space Agency (ESA) mission {\it Gaia} (\url{https://www.cosmos.esa.int/gaia}), processed by the {\it Gaia} Data Processing and Analysis Consortium (DPAC, \url{https://www.cosmos.esa.int/web/gaia/dpac/consortium}). Funding for the DPAC has been provided by national institutions, in particular the institutions participating in the {\it Gaia} Multilateral Agreement.

% % % % % %

\end{document}